\documentclass[journal=jacsat,manuscript=article]{achemso}

\usepackage[version=3]{mhchem} 
\usepackage[table]{xcolor} 

\usepackage[normalem]{ulem}

\author{Nicholas P. Sloane}
\altaffiliation{ARC Centre of Excellence in Exciton Science}
\author{Christopher G. Bailey}
\altaffiliation{ARC Centre of Excellence in Exciton Science}
\affiliation[UNSW]
{School of Physics, University of New South Wales, Sydney}
\author{Jared H. Cole}
\altaffiliation{ARC Centre of Excellence in Exciton Science}
\affiliation[RMIT]
{School of Science, Royal Melbourne Institute of Technology, Melbourne}
\author{Timothy W. Schmidt}
\altaffiliation{ARC Centre of Excellence in Exciton Science}
\affiliation[UNSW]
{School of Chemistry, University of New South Wales, Sydney}
\author{Dane R. McCamey}
\altaffiliation{ARC Centre of Excellence in Exciton Science}
\affiliation[UNSW]
{School of Physics, University of New South Wales, Sydney}
\author{Mykhailo V. Klymenko}
\affiliation{CSIRO's Data61}
\email{mike.klymenko@data61.csiro.au}
\title[An \textsf{achemso} demo]
  {Electronic Structure at the Perovskite/Rubrene Interface: The Effect of Surface Termination}

\begin{document}

\begin{tocentry}





\includegraphics[height = 4.15cm]{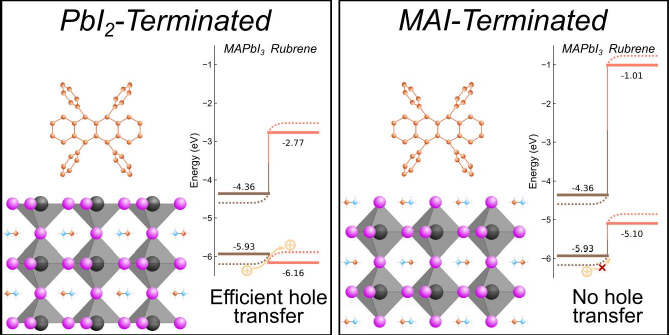}

\end{tocentry}

\begin{abstract}
    Perovskite films have rapidly emerged as leading active materials in optoelectronic devices due to their strong optical absorption, high carrier mobility and ease of fabrication. Whilst proving to be promising materials for solar cells and light-emitting diodes, another application of perovskites which makes effective use of their unique properties is sensitisation for photon upconversion. Consisting of a bulk perovskite sensitiser alongside an adjacent organic semiconductor film, the upconverting system can absorb multiple low-energy photons to emit high-energy photons. In this work, density functional theory, in conjunction with GW theory, is utilised to investigate the electronic structure at the MAPbI$_3$/rubrene interface for different surface terminations of MAPbI$_3$. From this investigation, we reveal that the surface termination of the perovskite layer greatly affects the charge density at the interface and within the rubrene layer driven by the formation of interfacial dipole layers. The formation of a strong interfacial dipole for the lead-iodide terminated perovskite alters the band alignment of the heterojunction and is expected to facilitate more efficient hole transfer, which should enhance triplet generation in rubrene through sequential charge transfer from the perovskite layer. The formation of this dipole layer is explained by the slight ionization of rubrene molecules due to the passivation of surface states. For the perovskite surface terminated with the methylammonium iodide layer, the highest occupied molecular orbital of the adjacent rubrene layer lies deep within the perovskite band gap. This termination type is further characterized by a lower density of states near the band edges and lower electron density, thereby acting as a spacer which is anticipated to decrease the probability of charge transfer across the interface. Thus based on our results, PbI$_2$-terminated perovskite surfaces are predicted to be favourable for applications where hole transfer to a rubrene layer is ideal, highlighting the significance of surface termination not only for upconverting systems but for all systems where the electronic environment at the interface is crucial to performance.
\end{abstract}

\section{Introduction}
Organometallic halide perovskites have rapidly risen to become leading candidates for use in devices such as photovoltaics\cite{Kojima2009OrganometalCells,Lee2012EfficientPerovskites,Green2014TheCells,Liu2023BimolecularlyCells} and light-emitting diodes.\cite{Tan2014BrightPerovskite,Lin2018PerovskiteCent,Ji2021HalideTechnologies} Perovskites possess properties well suited to such applications including solution processability, tunable bandgaps,\cite{Eperon2014FormamidiniumCells} high carrier diffusion lengths\cite{Ponseca2014OrganometalRecombination,Galkowski2016DeterminationSemiconductors} and high absorption cross sections.\cite{DeWolf2014OrganometallicPerformance,Bailey2019High-EnergySpectrophotometry} Three-dimensional organometallic halide perovskites follow the general formula ABX$_3$, where A is an organic cation such as methylammonium (CH$_3$NH$_3$), B is a metallic cation (commonly Pb), and X is a halide anion. While perovskites are established as excellent materials for use in optoelectronic devices, their properties also make them promising candidates for application as solid-state sensitisers for photon upconversion. 

Photon upconversion is a multi-photon process whereby a photon of higher energy is produced following the absorption of two or more lower energy photons in a suitable material. Upconversion shows considerable promise as a mechanism enabling the use of sub-bandgap photons for circumventing the detailed-balance limit\cite{Shockley1961DetailedCells} in single junction solar cells. One technique that permits photon upconversion is triplet-triplet annihilation (TTA), where two spin-1 triplet excitons can interact to produce one higher-energy, emissive, spin-0 singlet exciton. TTA upconversion has advantages over other upconversion pathways such as using lanthanide ions,\cite{Dong2013BasicEmissions} as it can be efficient at sub-solar fluences and with non-coherent excitation.\cite{Singh-Rachford2010PhotonAnnihilation,Schmidt2014PhotochemicalKinetics,Schulze2015PhotochemicalConversion,Alves2022ChallengesUpconversion,Feng2023PhotochemicalUpconversion} However, a consequence of utilising TTA for photon upconversion is that direct optical generation of triplet excitons are forbidden by spin selection laws.\cite{Dexter1953ASolids} To allow for the generation of triplet excitons, a typical TTA upconversion system is comprised of two main components, a sensitiser and an annihilator. The role of the sensitiser is to absorb incoming photons to generate singlet excitons, which undergo intersystem-crossing to a triplet exciton state. The resulting triplet excitons transfer to the annihilator where TTA can occur, generating singlet excitons that can recombine to emit a higher energy photon than those absorbed by the sensitiser.

In 2019 \citeauthor{Nienhaus2019Triplet-SensitizationUpconversion} demonstrated a new method of sensitised-TTA using a bulk 3D perovskite film as a sensitiser layer to generate triplet excitons in an adjacent film of rubrene.\cite{Nienhaus2019Triplet-SensitizationUpconversion,Wieghold2019TripletFluxes} Using a perovskite film as a solid-state sensitiser has a few advantages over alternate sensitisers by not relying on intersystem-crossing, thereby bypassing energy losses due to exchange energies in organometallic complexes\cite{Singh-Rachford2010PhotonAnnihilation} and by not being limited by poor exciton transfer seen in confined semiconductor sensitisers.\cite{Wu2016Solid-stateNanocrystals,Mase2017TripletUpconversion} Instead, due to the low exciton binding energy in perovskites, the triplet sensitisation mechanism was proposed to be due to asynchronous transfer of free charges from the perovskite into a bound triplet exciton in the rubrene, where holes could be injected into the rubrene HOMO (highest-occupied molecular orbital) from the perovskite valence band. Due to the large energy difference between the conduction band of the perovskite and the LUMO (lowest unoccupied molecular orbital) of rubrene, direct electron injection is not possible.\cite{Ji2017InterfacialInterface} It was instead proposed that, following hole transfer to rubrene, the electrons transfer to form triplet excitons.\cite{Nienhaus2019Triplet-SensitizationUpconversion,Wieghold2019TripletFluxes}

Questions still remain however over the exact sensitisation mechanism that occurs across the perovskite/rubrene heterojunction and whether triplet excitons in rubrene are formed via direct charge transfer, or via an intermediate state at the interface.\cite{VanOrman2021BulkUpconversionb} There have been multiple investigations using different fabrication techniques designed to alter the interface between the perovskite and rubrene layers to probe how the interfacial properties affect the triplet sensitisation process in these bilayers,\cite{Prashanthan2020InterdependenceAnnihilators,Wang2021InterfacialUpconversion,Sullivan2023SurfaceUpconversion} but they report conflicting findings. A major problem with perovskite materials is their sensitivity to many different variables that can be introduced throughout the fabrication and characterisation process, making it difficult to unravel the property responsible for an observed change.

In this work the electronic structure at a perovskite/rubrene interface is probed utilising a combination of density functional theory (DFT) and GW theory for two different surface terminations of methylammonium lead triiodide (MAPbI$_3$). The perovskite composition MAPbI$_3$ was chosen for the investigation as it has been shown to perform as solid-state sensitiser for upconversion with rubrene\cite{Prashanthan2020InterdependenceAnnihilators} and its relative structural/compositional simplicity as compared to other mixed cation/halide perovskites, reducing the complexity of the model. Previous theoretical studies have also shown how surface termination impacts properties of MAPbI$_3$ films such as; moisture degradation pathways,\cite{Mosconi2015AbWater,Koocher2015PolarizationSurfaces,Zhang2016TheInvestigation} thermodynamic stability\cite{Haruyama2014TerminationCells,Quarti2017InfluenceInterface} and most relevant for perovskite/rubrene bilayers, the electronic structure at the interface with organic semiconductors.\cite{Quarti2017InfluenceInterface} We find that the surface termination has a drastic effect on the charge redistribution at the MAPbI$_3$/rubrene interface, which is expected to have a large impact on the sensitisation of triplet excitons in rubrene following photoexcitation of the perovskite sensitiser. As a result, surface termination presents itself as a crucial variable that needs to be carefully controlled when energetic alignment is critical, not only for solid-state upconversion but also for light-emitting diodes and photovoltaic devices.

\begin{figure}
  \includegraphics[width = \linewidth]{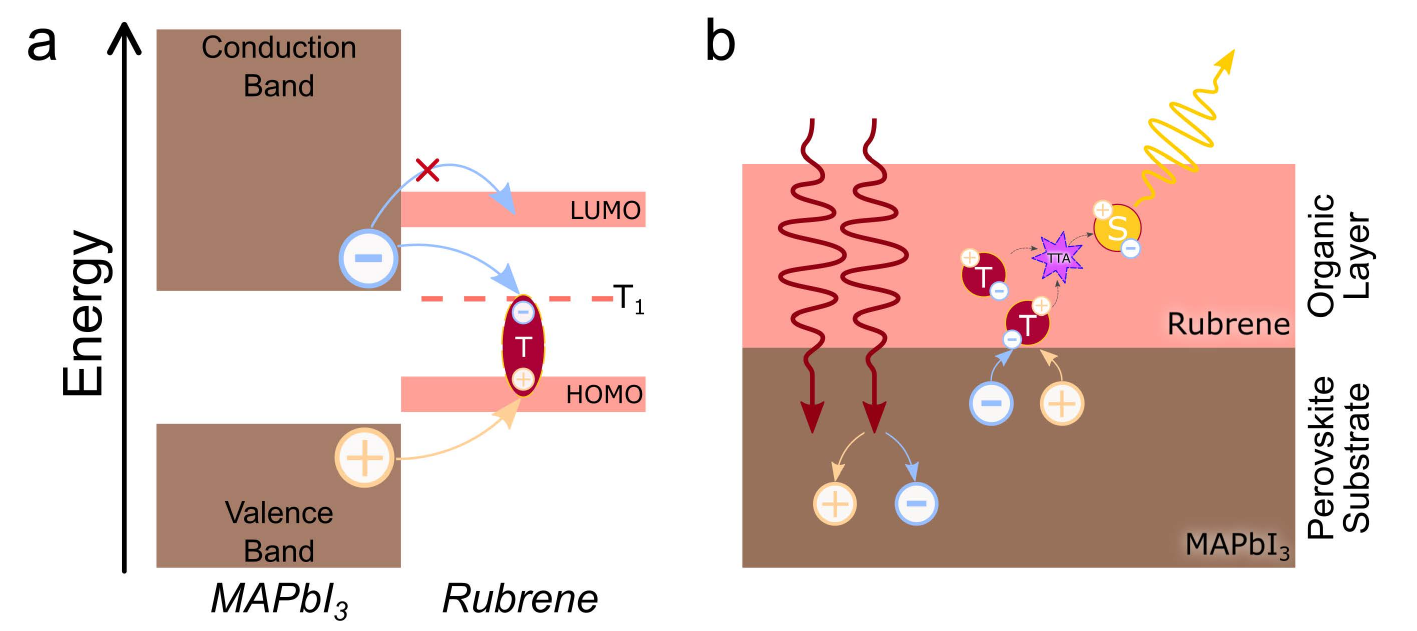}
  \caption{a) Energetic alignment between MAPbI$_3$ band edges and rubrene HOMO/LUMO. The alignment allows for hole transfer from the perovskite valence band to the rubrene HOMO but not for electron transfer to the rubrene LUMO. Instead electrons can transfer to a triplet exciton state in the rubrene (T$_1$) following hole transfer. b) Schematic outlining suggested upconversion process seen for perovskite/rubrene bilayers. Incoming near-infrared photons are absorbed by the perovskite which can sensitise triplet excitons in the rubrene. Two triplet excitons in the rubrene can interact to generate  singlet excitons capable of emitting a photon higher than those originally absorbed.}
  \label{fgr:bilayer}
\end{figure}

\section{Density Functional Theory Results}

\subsection{Atomic Structure of MAPbI$_3$/rubrene Supercells and DFT Computations}

\begin{figure}
    \centering
    \includegraphics[width=\linewidth]{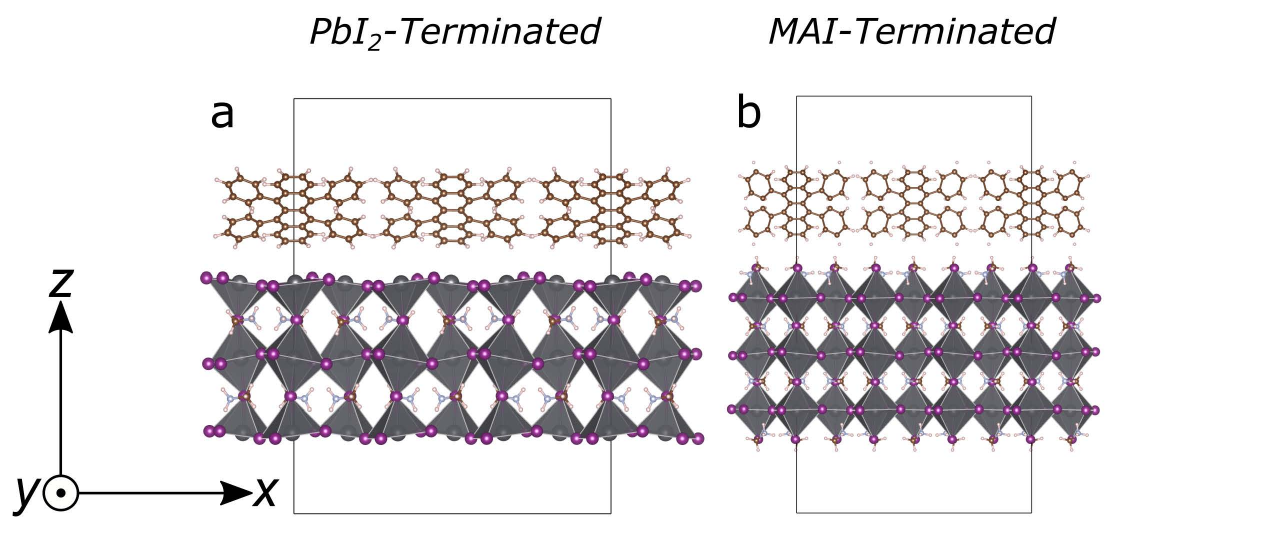}
    \caption{Relaxed atomic structures of the a) PbI$_2$-terminated and b) MAI-terminated supercells shown in the \textit{xz} plane. The boxes seen for each structure illustrates the size of the unit cells.}
    \textbf{\label{fig:geometry}}
\end{figure}

For the DFT computations, we have prepared two supercells, each consisting of a MAPbI$_3$\cite{Jain2013Commentary:Innovation} slab in contact with a monolayer of rubrene molecules,\cite{Jurchescu2006Low-temperatureTransport} but with different surface terminations for the perovskite slab. The two surface terminations are distinguished by whether the MAPbI$_3$ slab is terminated with a lead iodide layer (PbI$_2$-terminated) or with a methylammonium iodide layer (MAI-terminated). The MAPbI$_3$ slabs consist of three lead iodide layers with the methylammonium iodide layers varying based on the surface termination. For our model, we use the orthorhombic phase of MAPbI$_3$ for both terminations, since DFT computations have been performed at T = 0K, where the orthorhombic phase predominates.\cite{Poglitsch1987DynamicSpectroscopy,Piana2019Phonon-AssistedPerovskites} It is worth noting that the tetragonal phase of MAPbI$_3$ is more prevalent at room temperature. However, the results and conclusions of this paper are applicable to both phases. This is because both phases can exhibit either PbI$_2$- or MAI-terminations; the primary differences lie in slight deformations of the unit cell and the orientation of the organic MA cations. While the phases are characterised by different electron-phonon interactions,\cite{Saidi2018EffectsPerovskites} these variations are not expected to significantly impact the interface properties studied in this work. The discrepancies in atom positions within atomic layers between the two cases have a negligible impact on the results, which are predominantly determined by the in-plane average charge density and the density of states. The size of the unit cell for the geometry of each system was also increased in the \textit{z} direction so as to include a sufficiently wide vacuum layer to resolve the vacuum level which must be known to compute ionization energy and electron affinity. The relaxed structures are shown in \textbf{Figure \ref{fig:geometry}}. The DFT computations were performed using the plane-wave self-consistent field 
package of the Quantum Espresso suite.\cite{Giannozzi2009QUANTUMMaterials,Giannozzi2017AdvancedESPRESSO} The atomic co-ordinates were obtained by geometry optimisation with DFT using a plane-wave basis set and a norm-conservative pseudo-potentials retrieved from the PseudoDojo project.\cite{Hamann2013OptimizedPseudopotentials,vanSetten2018TheTable} The dispersion forces of the physisorbed rubrene monolayer on top of the MAPbI$_3$ are introduced into the model via the non-local exchange-correlation functional vdW-DF2-C09.\cite{CooperVanFunctional} The relaxation and self-consistent computations were performed on the 2 $\times$ 8 $\times$ 1 Monkhorst-Pack \textit{k}-space grid, using a kinetic energy cutoff of 80 Ry for wavefunctions and 320 Ry for charge densities. A denser \textit{k}-space grid was utilised for non-self consistent calculations performed on the 4 $\times$ 10 $\times$ 1 Monkhorst-Pack \textit{k}-space grid with the same kinetic energy cutoffs used. For all computations, the system is treated with periodic boundary conditions in the \textit{x} and \textit{y} directions. The 3D visualisations of atoms/electron density isosurfaces were generated using the VESTA (Visualization for Electron and STructural Analysis) program.\cite{Momma2008VESTA:Analysis}

\subsection{Termination-Dependent Dipole Layer}

Following the relaxation of both supercells with MAPbI$_3$ layers of different surface terminations, electrostatic potential and the charge density within the supercells were computed first using a self-consistent calculation. Alongside the supercells, the isolated slabs of MAPbI$_3$ and rubrene from both termination types were similarly computed, using the same geometries as in the heterostructure. The effect of the contact of two slabs can be estimated by subtracting the electrostatic potentials (V) and charge densities ($\rho$) of the isolated slabs from the corresponding characteristics of the slabs in contact. The change in electrostatic potential ($\Delta V$) and electron density ($\Delta \rho$) can be determined by:
\begin{equation}\label{delta V}
    \Delta V = V_{Per/Rub} - V_{Per} - V_{Rub}
\end{equation}
\begin{equation}\label{delta rho}
    \Delta \rho = \rho_{Per/Rub} - \rho_{Per} - \rho_{Rub}
\end{equation}

\begin{figure*}
    \centering
    \includegraphics[width=\linewidth]{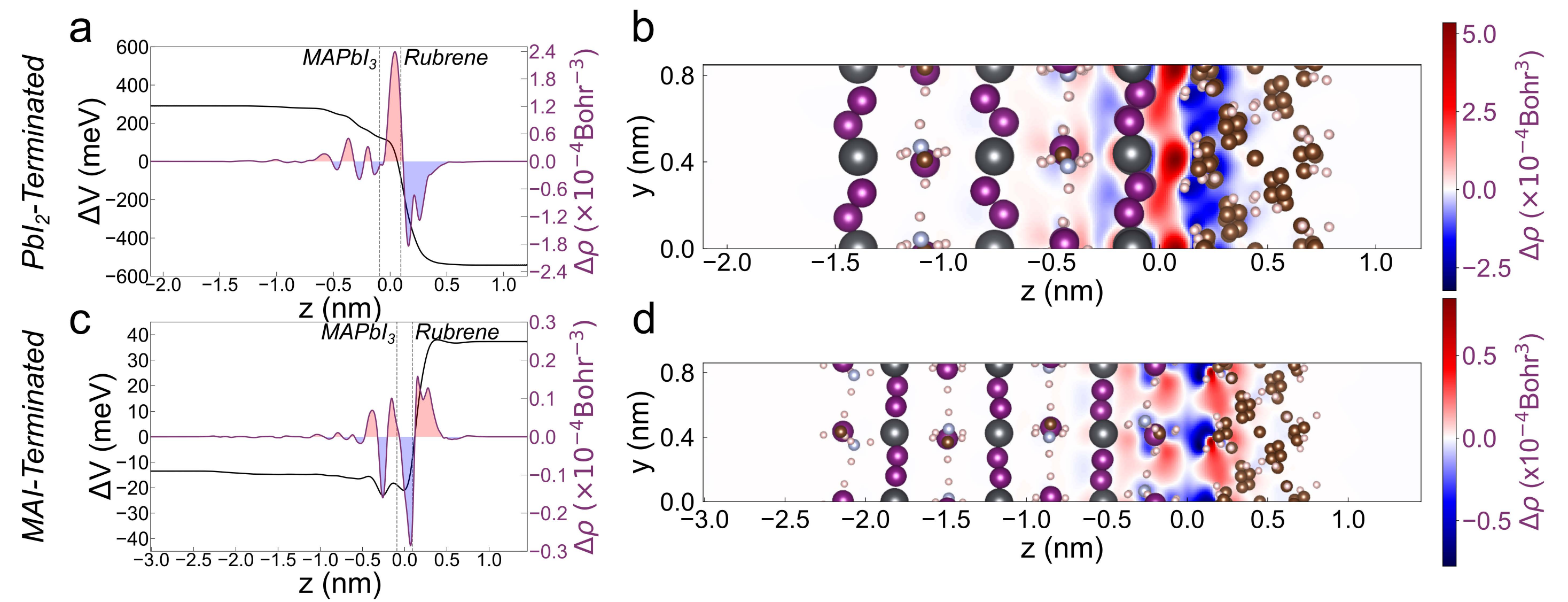}
    \caption{In-plane averaged electrostatic potential and charge density for a) the PbI$_2$-terminated and c) MAI-terminated supercells, where the dashed lines represent the interface. Also shown in b) and d) is a 2D plot of the difference in electron density due to the interaction of the slabs for the PbI$_2$- and the MAI-terminated systems respectively}
    \label{fig:V and rho}
\end{figure*}

Where the subscript Per/Rub describes the total supercell with the adjacent perovskite and rubrene slabs and the subscripts Per and Rub refer to the individual slabs of perovskite and rubrene computed for the comparison. The in-plane averaged electrostatic potential and charge density is shown for both PbI$_2$- and MAI-terminated systems in \textbf{Figures \ref{fig:V and rho}a} and \textbf{\ref{fig:V and rho}c} respectively.

\textbf{Figures \ref{fig:V and rho}a} and \textbf{c} show that the change in the electrostatic potential energy for both termination types is step-like in nature but opposite in sign to each other. A corresponding change in the density of electrons is also seen in \textbf{Figure \ref{fig:V and rho}} for both systems, which appears to indicate the presence of an interfacial dipole layer, as evidenced by the regions of higher and lower electron density resulting from the interaction of the slabs. The PbI$_2$-terminated system shows a strong increase of electron density at the MAPbI$_3$/rubrene interface and also demonstrates a clear reduction in the electron density in the within the rubrene layer. For the MAI-terminated supercell, a much weaker interaction is observed, indicating the formation of a less pronounced dipole and a correspondingly smaller change in electron density.

These results suggest that both systems possess opposite interfacial dipoles depending on the surface termination of the MAPbI$_3$ perovskite, with dipole energies of $\displaystyle E_{dip}$ = 830 meV for the PbI$_2$-terminated system and $\displaystyle E_{dip}$ = 50 meV for the MAI-terminated system. Notably, the dipole energy for the MAI-terminated supercell is an order of magnitude smaller than that of the PbI$_2$ supercell, suggesting that the interaction between the perovskite and rubrene slabs is significantly weaker in the MAI-terminated system. The dipole energy is defined as the difference in the vacuum energy at either end of the \textit{z} axis of the supercells (\textbf{Figure \ref{fig:geometry}}). 

\begin{figure}
    \centering
    \includegraphics[width=\linewidth]{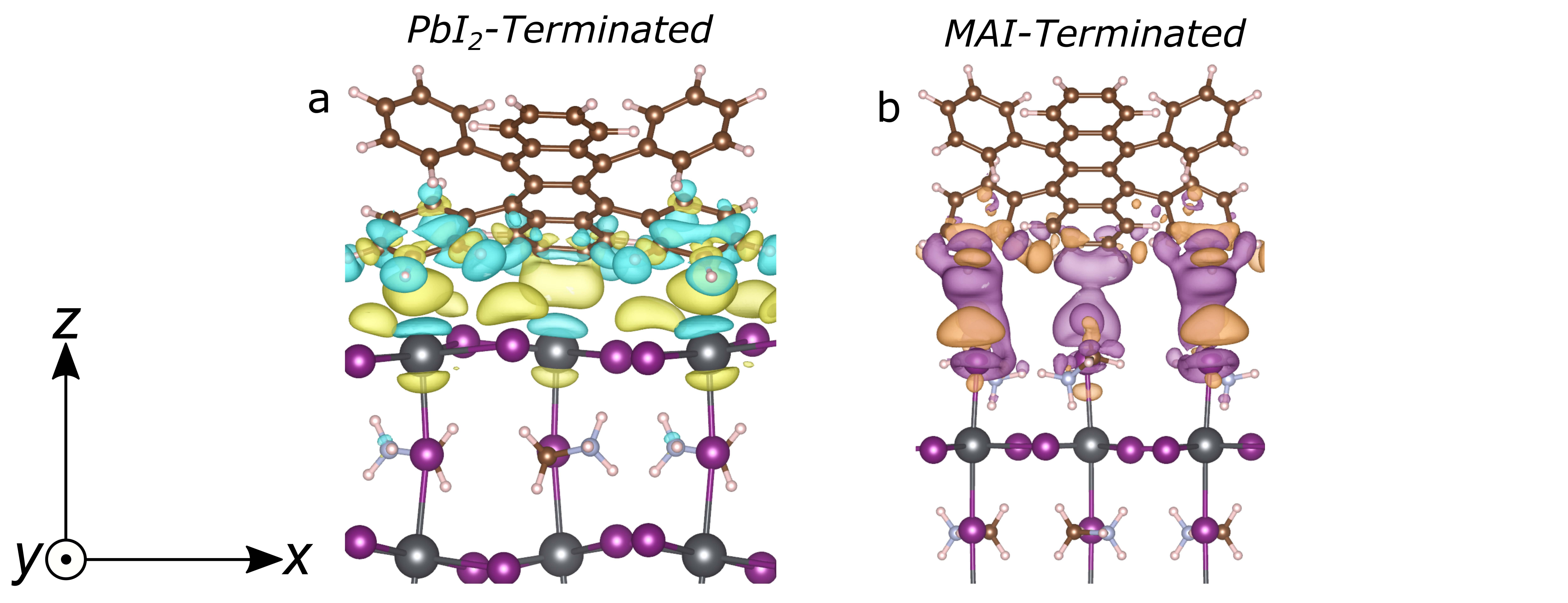}
    \caption{Charge difference for the a) PbI$_2$-terminated and b) MAI-terminated supercells. a) Electron gain/loss is shown in yellow/cyan with isosurface value $5\times10^{-4}$ electron/Bohr$^{3}$. b) Electron gain/loss is shown in orange/blue with isosurface value $1\times10^{-4}$ electron/Bohr$^{3}$.}
    \textbf{\label{fig:rho comp}}
\end{figure}

The origin of these dipoles is critical to understand as they are expected to greatly affect the overall behaviour and mechanisms in real-world interfaces. Dipoles can be produced by various mechanisms such as direct charge transfer from one layer to the other, orbital hybridisation, or from the rearrangement of electrons. The specific origins of the dipole layer for each type of termination are further discussed in the following sections. The change in the electron density due to the interaction of the slabs is shown in \textbf{Figure \ref{fig:rho comp}}, with relation to the atomic positions. In the PbI$_2$-terminated system (\textbf{Figure \ref{fig:rho comp}a}) there is a clear region of increased electron density immediately adjacent to the perovskite layer, and a region of decreased electron density in the rubrene layer overall. However, in \textbf{Figure \ref{fig:rho comp}b} it can be seen that the charge density redistribution is of a different nature in the MAI-terminated system compared to the PbI$_2$ system. Overall there appears to be a reduction in the density of electrons across the interface with the redistribution of the electrons occurring in the organic methylammonium groups and the rubrene layer.  The isosurface value shown for the MAI-terminated system in \textbf{Figure \ref{fig:rho comp}b} is one fifth of that shown in \textbf{Figure \ref{fig:rho comp}a}, as the total change in charge density in the MAI-terminated system is far smaller than that seen for the PbI$_2$ system.

Thus far, the results point towards surface termination of MAPbI$_3$ greatly affecting the electronic interaction with the adjacent rubrene layer, in agreement with a study reported by \citeauthor{Quarti2017InfluenceInterface}\cite{Quarti2017InfluenceInterface} for a system consisting of MAPbI$_3$ and C60. Conversely, \citeauthor{Quarti2017InfluenceInterface} found that a surface terminated with MAI was beneficial for electron transport to an adjacent C60 molecule. Our results indicate the formation of a dipole layer at the interface between MAPbI$_3$ and rubrene, with a strong dipole forming at the interface in the PbI$_2$-terminated system and a weaker interaction in the MAI-terminated system, with redistribution of charges within each slab being the dominant effect.

The depletion of electrons at the interface with MAI-termination can be explained by the Pauli pushback effect, where charges are redistributed to reduce the overlap of orbitals between the slabs as a result of contact.\cite{Zojer2019TheInterfaces} However, the accumulation of electrons at the PbI$_2$ interface requires further analysis, which will be carried out in the following sections.

\subsection{Local Density of States}

\begin{figure}
    \centering
    \includegraphics[width=\linewidth]{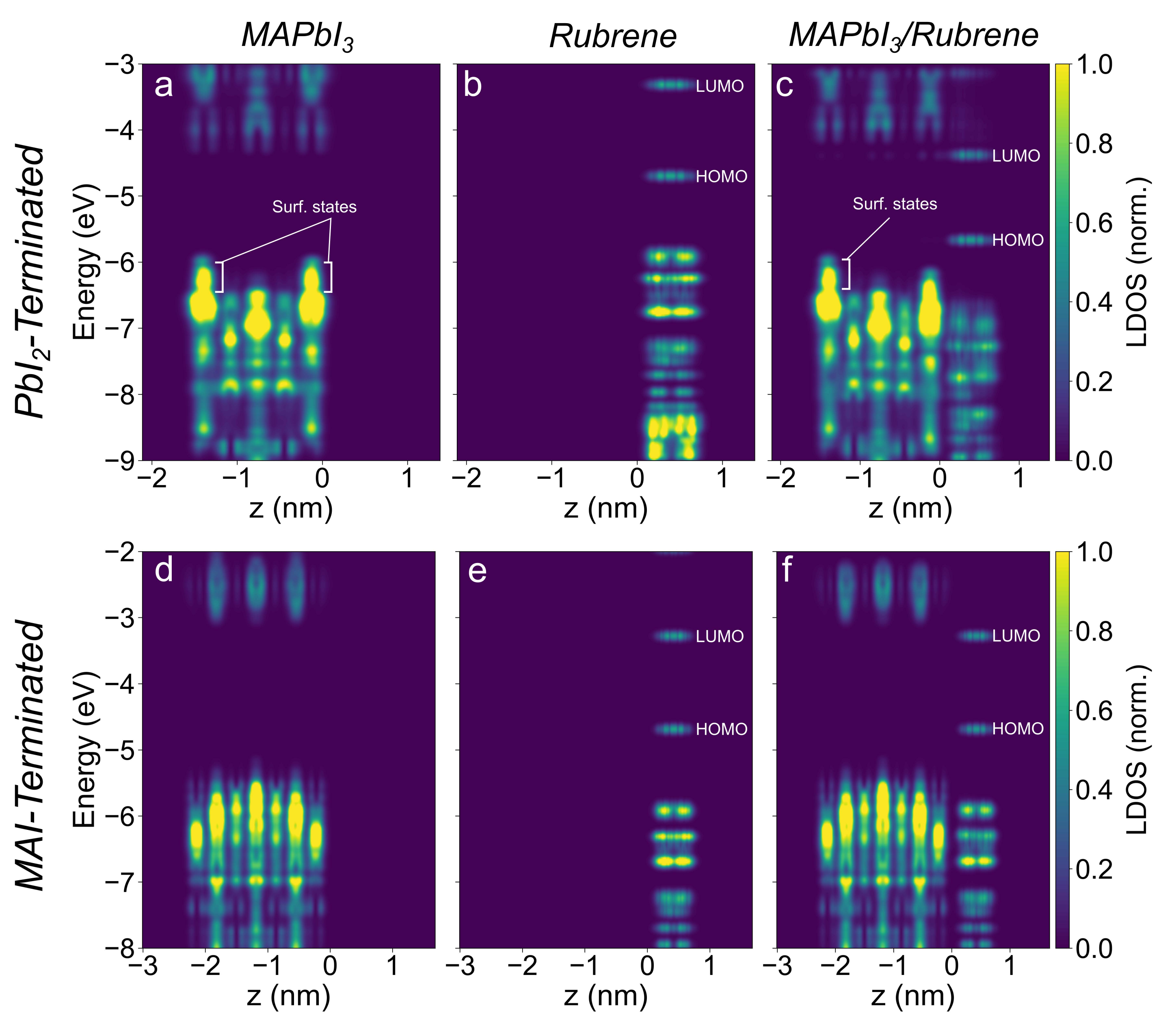}
    \caption{Normalised Local Density of States averaged along the \textit{xy} plane. Surface states in the PbI$_2$-terminated perovskite are outlined alongside the energetic position of the HOMO and LUMO in the rubrene. Maximum and minimum values are truncated for visibility.}
    \label{fig:LDOS}
\end{figure}

So far we have demonstrated how the interaction of the perovskite and rubrene slabs impact the electrostatic properties of the total system. Another way to visualise the interaction both spatially and energetically is via observing the local density of states (LDOS) for each system. The LDOS for both the PbI$_2$- and MAI-terminated supercells are shown in \textbf{Figure \ref{fig:LDOS}}, where the LDOS is a function of both energy and the spatial coordinate \textit{z}, which is perpendicular to the interface between the slabs. The atomic contributions to the density of states is shown in \textbf{Figure \ref{fig:supp_pdos}}. From \textbf{Figure \ref{fig:LDOS}a} and \textbf{d} it can be seen that the LDOS within the perovskite slab ($z<0$) has clear layers alternating between regions of high DOS and regions of lower DOS. Comparing the different types of termination we can determine that the dense regions originate from the PbI$_2$ layers and the lower density regions to be a result of the MAI layers based on which layers are present on the surface of the respective supercells. Two other observations can also be made: i) the position of the valence band maximum (VBM) and the conduction band minimum (CBM) shifts depending on the termination type of the perovskite and ii) the relative HOMO and LUMO alignment of the rubrene layers ($z>0$) also shifts depending on surface of the perovskite. These results add further evidence to suggest that the termination type strongly affects the interaction of the rubrene and perovskite slabs. The stronger interaction for the PbI$_2$-terminated perovskite with the rubrene slab proposed in the previous section can potentially be explained by the adjacent PbI$_2$ layer acting as dense reservoir of states to interact with. For the MAI-terminated perovskite however, the lower density MAI layers potentially act as spacers or barriers between the PbI$_2$ layers resulting in far weaker overall interaction between the rubrene and MAPbI$_3$ slab.

Additionally, in the case of PbI$_2$-termination, we observe the formation of additional surface states slightly above the valence band edge in perovskite. By computing the electron localisation function for the two supercells (Figure \ref{fig:supp_elf}), these surface states can be determined to be defect states and not dangling bonds due to electron density being localised at the iodine atomic centres. Further, the adjacent rubrene monolayer appears to partially passivate these surface states as seen in \textbf{Figure \ref{fig:LDOS}c} where the tip of the surface states shifts down in energy when in contact in rubrene. To estimate whether the proposed interfacial dipoles are caused by charge transfer or by other processes such as orbital hybridisation, the change in the LDOS can be calculated as a result of the interaction of the slabs. Following from Equation (\ref{delta rho}) the change in charge density can be described also by:
\begin{equation}
\label{Delta rho = LDOS}
\centering
    \Delta \rho = \int_{-\infty}^{E_{f}^{Per/Rub}} \Delta LDOS(\epsilon) d\epsilon
\end{equation}
Where $E_{f}^{Per/Rub}$ is the computed Fermi energy of the total supercell and $\Delta LDOS(\epsilon)$ is described by \textbf{Equation (\ref{Delta LDOS})}:
\begin{align}
    \label{Delta LDOS}
    \Delta LDOS(\epsilon) &= LDOS_{Per/Rub}(\epsilon) \notag \\
    &- LDOS_{Per}(\epsilon - E_f^{Per/Rub} + E_f^{Per}) \notag \\ 
    &- LDOS_{Rub}(\epsilon - E_f^{Per/Rub} + E_f^{Rub})
\end{align}

Where similarly $E_f^{Per}$ and $E_f^{Rub}$ are the computed Fermi energies of the perovskite and rubrene slabs respectively.

The difference in LDOS is calculated by subtracting the LDOS of the isolated slabs from the LDOS of the total supercell after first aligning them by the Fermi energy ($E_f$) of each slab, which is defined as the onset of the valence band. A similar method was utilised for a supercell consisting of silicon and tetracene slabs, where the interaction between the inorganic and organic system was determined to be from orbital hybridisation.\cite{Klymenko2022AInterface} The difference in LDOS as a function of both energy and the z coordinate is shown in \textbf{Figure \ref{Delta LDOS}} for both MAPbI$_3$ terminations. Comparing the differences in the local density of states in the two termination types, there is a clear region of change close to the bandgap (~-4.3 to -6.8 eV MAPbI$_3$, -4.4 to -5.6 eV Rubrene) for the PbI$_2$-terminated supercell, whereas the changes are small in the MAI-terminated supercell close to the band gap energies (~-3.2 to -5.1 eV MAPbI$_3$, -3.3 to -4.7 eV Rubrene).

\begin{figure}
    \centering
    \includegraphics[width=\linewidth]{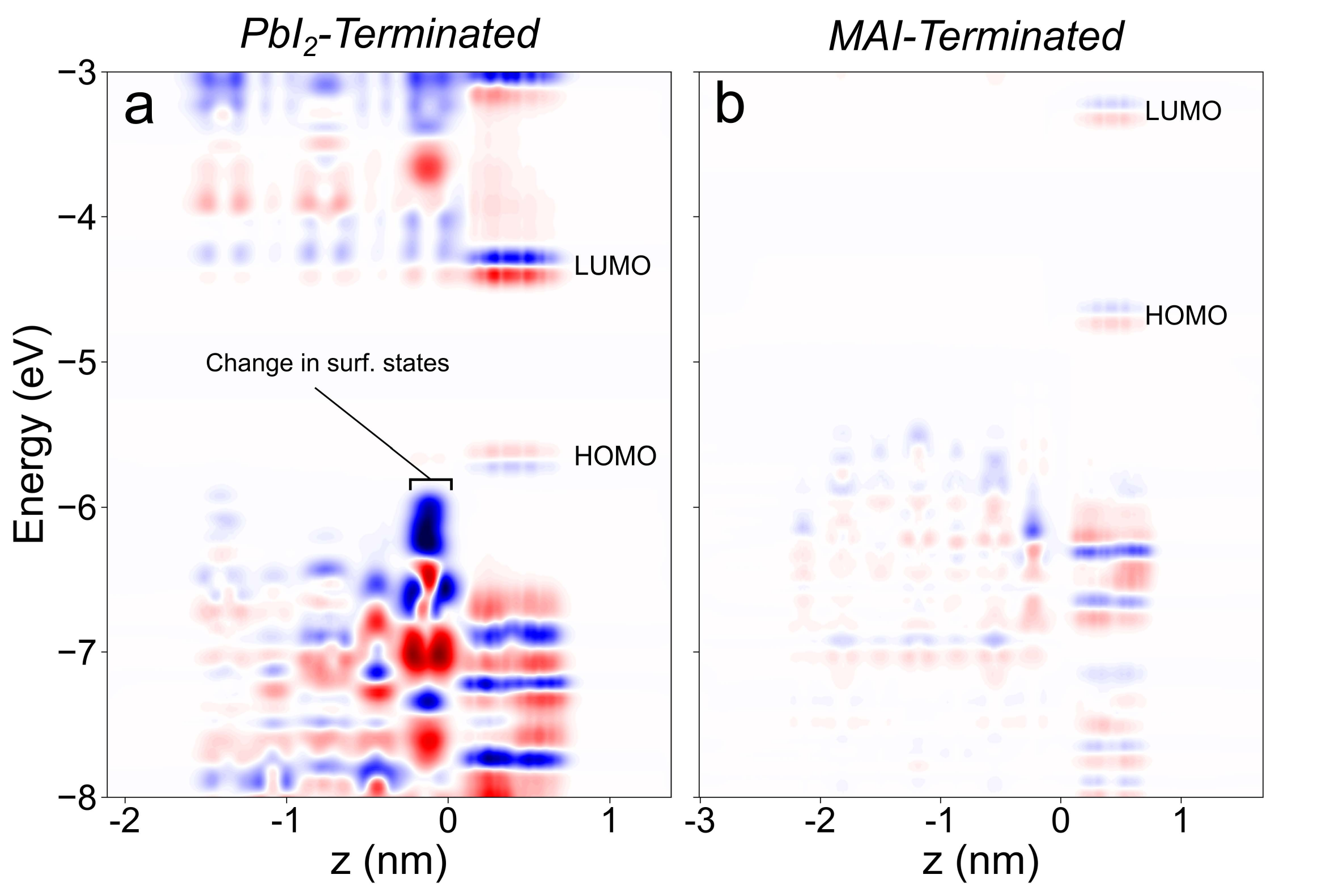}
    \caption{Difference in the local density of states as a result of the interaction between the MAPbI$_3$ and rubrene slabs for both the a) PbI$_2$-terminated and b) MAI-terminated perovskites. An increase in the LDOS is represented by the red regions and a decrease is represented by the blue regions.}
    \textbf{\label{fig:Delta LDOS}}
\end{figure}

The significant change in LDOS at the surface of perovskite in contact with organic molecules in the PbI$_2$-terminated system indicates the disappearance of surface states and a strong redistribution of electron density at the interface. In \textbf{Figure \ref{fig:Delta LDOS}a}, this disappearance is indicated by a large region of change at -6 eV at the MAPbI$_3$/rubrene interface. This is a result of the passivation of surface states by the organic molecules. As a result of passivating surface states, rubrene molecules acquire a slight positive charge, as confirmed by the results in \textbf{Figure \ref{fig:rho comp}a}.

In contrast, the small change in the density of states overall in the MAI-terminated system, as shown in \textbf{Figure \ref{Delta LDOS}b}, indicates almost no significant charge redistribution resulting from the contact between the perovskite and rubrene slabs.

\section{GW computations and energy bands alignment}
\subsection{GW Introduction}

Static DFT is fundamentally a ground state theory, and attempts to use it for predicting excited state properties, such as band gap energy, often result in significant underestimations.\cite{Perdew1985DensityProblem} In order to improve the accuracy of predictions related to excited states, we use many-body theory based on Hedin's equations and the GW approximation.\cite{Hedin1999OnApproximation} Utilising the Kohn-Sham orbitals (the basis set calculated by DFT) generated from the previous section, GW computations were performed using the BerkeleyGW code.\cite{Deslippe2012BerkeleyGW:Nanostructures} The computational procedure consists of two steps: first, the dielectric matrix ($\epsilon$) and its inverse ($\epsilon^{-1}$) are calculated using the Kohn-Sham orbitals as the basis set; second, the GW self-energy functions and corresponding corrections to quasi-particle energies are computed.\cite{Hedin1970EffectsSolids,Hedin1999OnApproximation,Onida2002ElectronicApproaches} Given the large sizes of the supercells and the high number of atoms, we employ the non-iterative \textit{G$_0$W$_0$} approximation to reduce the computational resources required.

\subsection{GW Results}

\begin{figure}
    \centering
    \includegraphics[width=\linewidth]{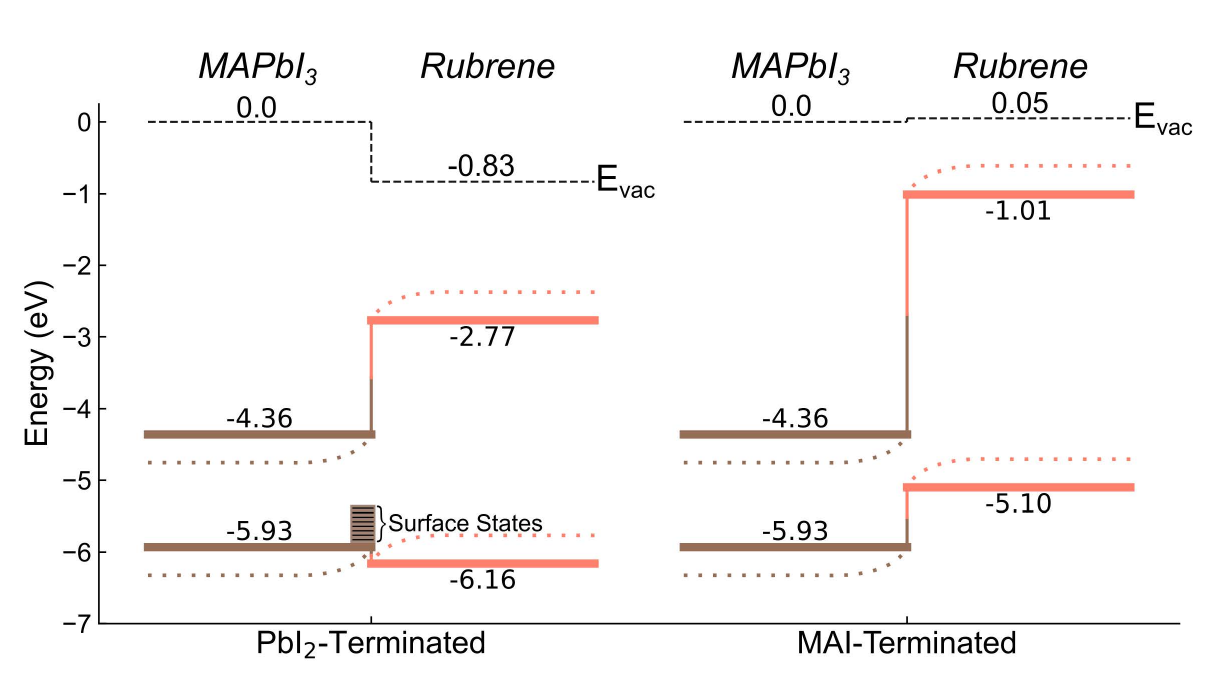}
    \caption{Energy band alignment at the MAPbI$_3$-rubrene interface for the a) PbI$_2$-terminated and b) MAI-terminated perovskites, computed using GW theory. The dotted lines indicate the effect of band bending between the  slightly n-doped perovskite the rubrene (a p-type organic semiconductor). The variation of the vacuum level, E$_{\text{vac}}$, across the interface indicates the presence of the interfacial dipole layer.}
    \label{fig:alignment}
\end{figure}

The results of the computations, based on the GW theory, for the band alignment are shown in \textbf{Figure \ref{fig:alignment}}. In this figure, we show the computed results for the rubrene only, while the band edges for the perovskite are those reported by \citeauthor{Luo2021RecentDevices}\cite{Luo2021RecentDevices} This approach has been implemented because accurately predicting band edges for perovskite using the GW method necessitates accounting for spin-orbit coupling, which effectively doubles the number of basis functions and significantly increases computational costs, rendering the problem intractable for large supercells. Furthermore, even when implemented, the slab model for inorganic semiconductors remains susceptible to errors due to the quantum confinement effect. This effect is comparable to the influence of rubrene on perovskite in terms of energy when the slab thickness is less than five nanometers.

The obtained results confirm the crucial role of surface termination: the PbI$_2$-termination shifts the HOMO levels of rubrene closer to the perovskite band edge, whereas for the MAI-termination, these levels lie within the perovskite band gap. The band offset for the PbI$_2$-termination is similar to that reported by \citeauthor{Ji2017InterfacialInterface}\cite{Ji2017InterfacialInterface} who reported that the HOMO band edge of rubrene was 100 meV below the perovskite band edge, while our results show it to be approximately 200 meV lower. This small discrepancy can be attributed to the fact that our computations were performed for a molecular monolayer, whereas the referenced study considered a bulk rubrene film. Although the observed band offset does not appear to favor hole transfer, considering the band bending caused by the doping of the semiconductors can make hole transfer possible (see the dotted lines in \textbf{Figure \ref{fig:alignment}} and discussion by \citeauthor{Ji2017InterfacialInterface}\cite{Ji2017InterfacialInterface}). For the MAI-terminated surface, the HOMO band of rubrene lies deep within the band gap of perovskite. Given that the frontier bands for organic semiconductors have small widths (several hundred meV), hole transport is expected to be highly unfavourable in this case.

Next, we compare the GW results for the supercells containing slab models of the rubrene/perovskite interfaces with both MAI- and PbI$_2$-terminated surfaces to those of isolated rubrene slabs to estimate the effect of the substrate on the electronic structure of the molecular crystal. To isolate the dynamic dielectric screening effects from those caused by crystal deformations resulting from the contact between the two materials, the geometry of the isolated rubrene slabs was obtained by first relaxing the rubrene on one of the two types of perovskite substrates (with either MAI- or PbI$_2$-terminations), then freezing the geometry of the molecules and removing the substrate. The calculated bandgaps for the isolated rubrene slabs are 4.16 and 4.41 eV for the slabs relaxed on the PbI$_2$- and MAI-terminated surfaces, respectively, shown in \textbf{Table \ref{tbl:table_eg}}. The slight difference in the bandgaps can be attributed to the minor variations in the geometry of the molecules in the monolayer, which result from their relaxation on different substrates. Note that the experimental HOMO-LUMO gap of bulk rubrene is reported to be ~2.9 eV\cite{Ji2017InterfacialInterface} (with an optical HOMO-LUMO gap corresponding to ~2.3eV),\cite{Irkhin2012AbsorptionCrystals} and this discrepancy is due to the reduced intermolecular interactions in the monolayer compared to the densely packed molecules in bulk crystals. The GW calculations of the HOMO-LUMO gap for an isolated rubrene molecule yield a value of 4.31 eV \cite{Wang2016EffectPolymorphs} using the G$_0$W$_0$ approximation and the basis set obtained from DFT with the PBE functional. Thus, a monolayer of molecules more closely resembles an isolated molecule in a vacuum than a molecule packed in a 3D crystalline structure, particularly in terms of dielectric screening and long-range correlation effects. Indeed, our GW calculations, using the same parameters as before for a bulk rubrene system with a relaxed orthorhombic structure shown in \textbf{Figure \ref{fig:supp_rubrene-geometry}}, yield a HOMO-LUMO gap of 2.53 eV. This value is much closer to the experimental value, confirming that the discrepancy was due to the lack of intermolecular interactions between the rubrene molecules in the monolayers and not due to the model itself.

The slab model with GW calculations is useful for determining the role of dielectric screening caused by the perovskite substrate on the rubrene energy levels. As outlined above, the bandgaps of the isolated rubrene slabs relaxed on the PbI$_2$ and MAI perovskite slabs were 4.16 and 4.41 eV respectively. We compare these numbers to the results obtained when the rubrene slab is in contact with perovskite. First, for the MAI-terminated substrate, the bandgap of the rubrene layer undergoes a downwards shift from 4.41 eV to 4.10 eV. The small decrease in bandgap energy of 0.31 eV suggests a limited effect that is purely caused by the dielectric screening and long-range electron correlations. Subsequently the same comparison was performed for the PbI$_2$-terminated supercell and a much stronger narrowing of the rubrene bandgap was observed, going from 4.16 eV for the isolated slab to 3.39 eV for the rubrene in the overall supercell. This narrowing of the rubrene band gap of ~0.8 eV indicates a much stronger interaction with the PbI$_2$-terminated perovskite substrate than that of the MAI-terminated perovskite.

Note that in the case of MAI-termination, the contact leads to an downward shift of both the IP and EA by 0.72 eV and 0.41 eV, respectively. In contrast, for the PbI$_2$-termination, the contact results in a decrease of IP by 0.62 eV and an increase of EA by 0.16 eV.

\begin{table}
    \caption{Computed Ionisation Potential (IP), Electron Affinity (EA) and Quasiparticle Energies (E$_g$) of rubrene for both PbI$_2$ and MAI supercells, compared to the rubrene slabs relaxed on the PbI$_2$ and MAI surfaces. The values for IP and EA were determined based on the difference of energies between the HOMO/LUMO and the vacuum energy, respectively. Bulk rubrene only has values for E$_g$ since there is no vacuum for comparison, due to the unit cell having repeating boundaries in all directions.}
  \label{tbl:table_eg}
  \begin{tabular}{|c|c|c|c|c|}
    \hline
    \rowcolor{gray!20} \multicolumn{2}{|c|}{\textbf{Type}} & \textbf{IP (eV)} & \textbf{EA (eV)} & \textbf{E$_g$ (eV)} \\
    \hline
    Organic Layer & Substrate Termination & & & \\
    \hline
    \rowcolor{gray!20} Isolated Rubrene & MAI & -5.83 & -1.42 & \textbf{4.41} \\
    \hline
    \rowcolor{gray!20} Rubrene (in contact) & MAI & -5.11 & -1.01 & \textbf{4.10} \\ \hline
    Isolated Rubrene & PbI$_2$ & -5.80 & -1.64 & \textbf{4.16} \\
    \hline
    Rubrene (in contact) & PbI$_2$ & -5.18 & -1.79 & \textbf{3.39} \\  \hline
    \rowcolor{gray!20} Bulk Rubrene & & &  & \textbf{2.53} \\
    \hline
  \end{tabular}
\end{table}

\section{Discussion and Conclusions}

Based on the computational results shown in the previous sections, it is clear that controlling the surface termination of the perovskite film plays a vital role in the electrostatics within MAPbI$_3$/rubrene bilayers. Due to the crucial role that surface termination appears to have on real-world devices and systems, it is an essential property to control during the fabrication process of perovskite films. Previous experimental studies found that increasing the relative amount of PbI$_2$ to MAI precursors leads to the transition from p-type to n-type doped films,\cite{Wang2014QualifyingCH3NH3PbI3} verifying previous DFT results reporting self-doping being determined by the stoichiometry of the films.\cite{Haruyama2014TerminationCells,Kim2014ThePerovskite,Yin2014UnusualAbsorber} Additionally, annealing MAPbI$_3$ films has been utilised to remove MAI from the surface of films due to the volatilisation of MAI at temperatures of 100$^\circ$C\cite{ Wang2017ScalingFilms,Lu2021InPathway} resulting in n-type doping of the resultant films.\cite{ Wang2014QualifyingCH3NH3PbI3} Furthermore, post-fabrication solvent treatment steps with polar solvents, \citeauthor{Wang2014QualifyingCH3NH3PbI3} treated perovskite films with isopropyl alcohol (polar) to dissolve trace organic halides from the surface of the perovskite films.\cite{Wang2021InterfacialUpconversion} \citeauthor{Sullivan2023SurfaceUpconversion} further observed that polar solvent treatment shifts the Fermi level towards the conduction band, indicating an n-type surface, potentially due to the formation of a PbI$_2$-terminated surface through the dissolution of MAI by the polar solvents.\cite{Sullivan2023SurfaceUpconversion}. 

The results of this computational investigation into the interfacial interaction between MAPbI$_3$ and rubrene show the important role that the perovskite surface termination has on the electronic structure between the two layers. For the PbI$_2$-terminated supercell there was a strong interaction between the MAPbI$_3$ and the rubrene due to the greater DOS for the PbI$_2$ layers, resulting in the formation of a charge-transfer dipole. The dipole can be interpreted from the electron density shown in \textbf{Figure \ref{fig:rho comp}} to be an accumulation of electron density at the interface and a corresponding reduction of electron density in the rubrene. The formation of this dipole layer and slight ionization of rubrene molecules are attributed to the passivation of surface states in the perovskite by rubrene when the former has the PbI$_2$-terminated surface. Surface states play a dual role: they partially attract charge from rubrene, contributing to the formation of a dipole layer, and they provide an additional pathway for hole transport across the heterojunction. While these surface states are passivated, they do not completely vanish; they remain present, albeit to a lesser extent on the energy scale compared to that without passivation (see \textbf{Figure \ref{fig:LDOS}a and c}). As is indicated by the computed band alignment, the dipole layer is expected to facilitate hole transfer by lowering the HOMO band of rubrene close to the valence band edge of perovskite (see \textbf{Figure \ref{fig:alignment}}). This, in turn, should facilitate triplet exciton sensitisation through the proposed sequential charge transfer mechanism.\cite{Nienhaus2019Triplet-SensitizationUpconversion,Wieghold2019TripletFluxes} Specifically, rubrene precharged with holes enhances the electron transfer from the perovskite conduction band into a triplet exciton in the rubrene.\cite{Wieghold2020PrechargingDevices}
The strong dipole layer should allow for the more rapid population of triplet excitons in rubrene due to its existing hole population as a result of the electronic structure formed across the heterojunction, leading to rapid TTA at the interface.\cite{Wieghold2019InfluenceUpconversion} 

Contrastingly, the MAI-terminated supercell demonstrated a much weaker interaction between the perovskite and the rubrene slabs due to the corresponding band alignment and the fact that the MAI layers possess lower overall electron density and density of states close to the band edges. Whilst there exists a rearrangement of charges in the system as a result of the interaction of the slabs, there was little reasoning to prove that this was a dipole layer as a result of charge transfer between the slabs and can rather be potentially attributed to the Pauli pushback/cushion effect.\cite{Zojer2019TheInterfaces} For the MAI-termination, although the valence band edge of rubrene is higher than that of the perovskite, hole transfer is not guaranteed due to the narrow width of the HOMO band. Perovskite surfaces terminated with MAI would thus serve as a spacer for charge or exciton transfer.

Thus based on our results, for applications where hole transfer is ideal from a perovskite to a rubrene layer PbI$_2$-terminated perovskite surfaces would be favourable. This is due to to the intrinsic charge-transfer dipole that is formed from the electronic structure between the layers, contrasted to MAI-terminated surfaces where there is much weaker/negligible interaction. Overall, surface termination presents itself as a crucial variable to control during the fabrication of perovskite films not only to select beneficial interfacial electronic structures for more effective charge transport within light-emitting diodes/solar cells but also for novel applications such as perovskite sensitised solid state upconversion.\cite{Nienhaus2019Triplet-SensitizationUpconversion,Wieghold2019TripletFluxes}

\begin{acknowledgement}

The authors acknowledge support through the Australian Research Council Centre of Excellence in Exciton Science (CE170100026). N.P.S acknowledges the support from an Australian Government Research Training Program (RTP) Scholarship. This research was undertaken with the use of the National Computational Infrastructure (NCI Australia). NCI Australia is enabled by the National Collaborative Research Infrastructure Strategy (NCRIS).

\end{acknowledgement}
\bibliography{references}
\begin{suppinfo}

\setcounter{figure}{0}
\makeatletter 
\renewcommand{\thefigure}{S\@arabic\c@figure}
\makeatother

\begin{figure}[H]
    \centering
    \includegraphics[width=\linewidth]{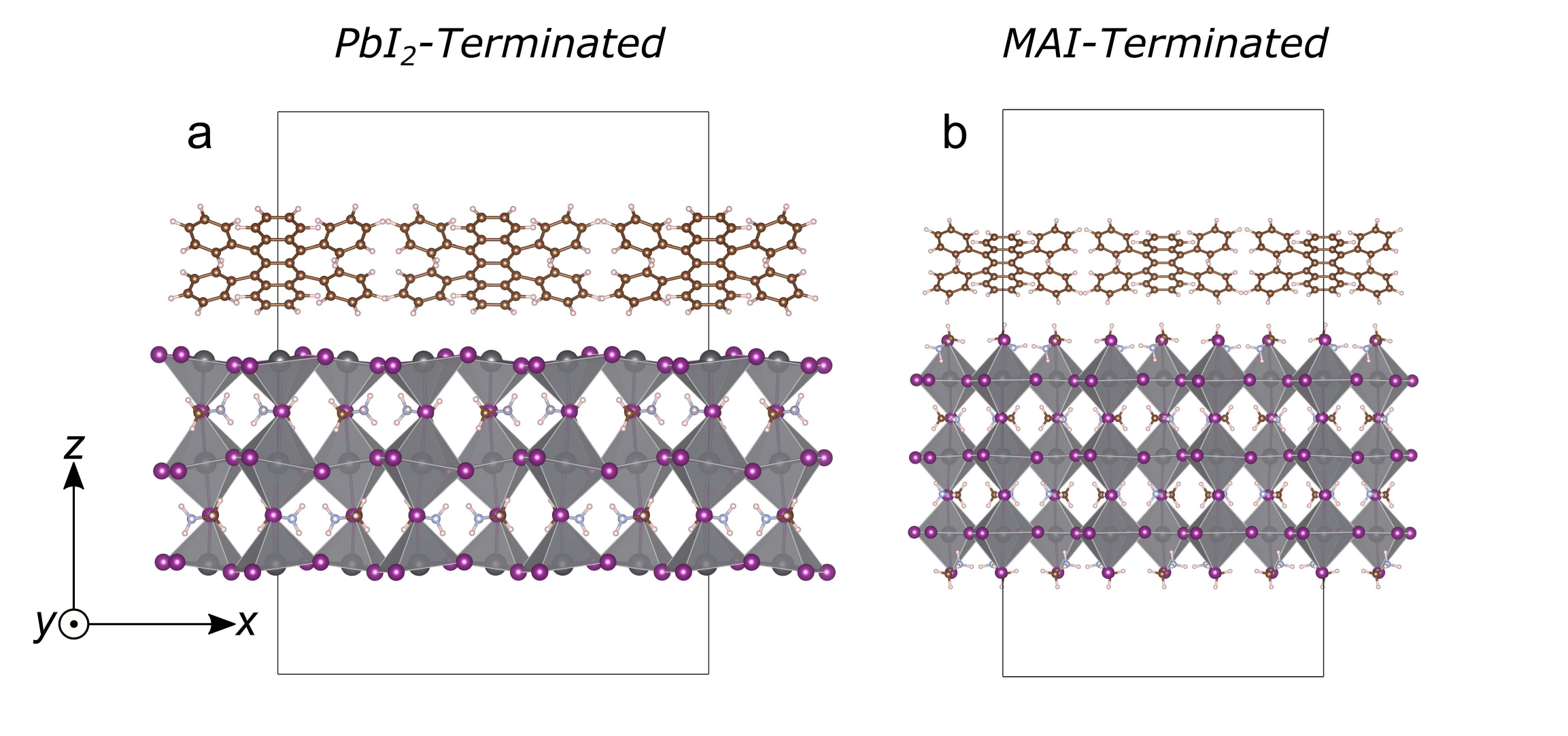}
    \caption{Initial geometries of a) PbI$_2$-terminated and b) MAI-terminated supercells prior to relaxation}
    \label{fig:supp_initial-geometry}
\end{figure}

\begin{figure}[H]
    \centering
    \includegraphics[width=\linewidth]{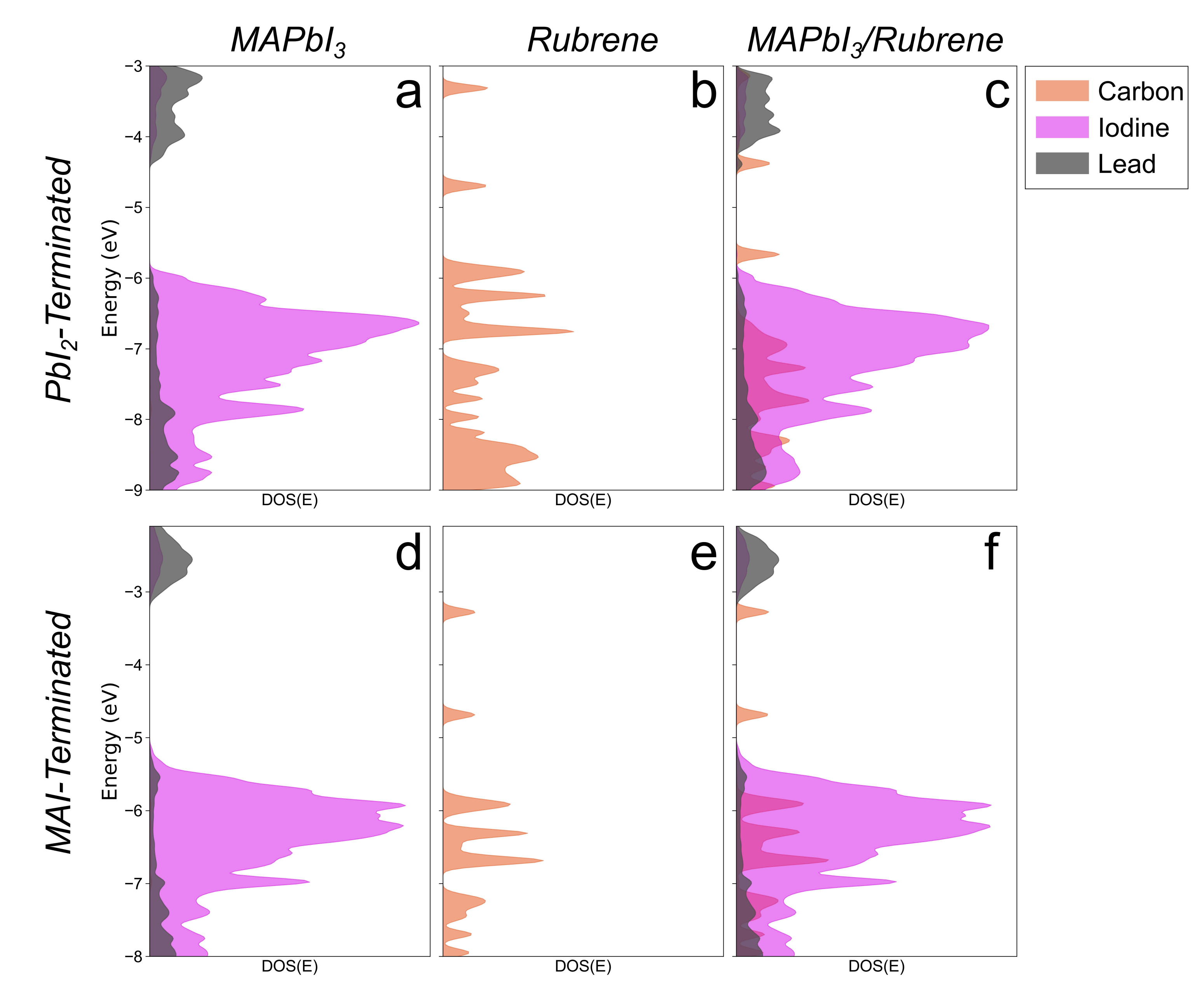}
    \caption{Partial Density of States (PDOS) for both termination types of slabs/supercells dictating the role played by Carbon, Iodine and Lead orbitals making up the Valence and Conduction band edges. Hydrogen and Nitrogen PDOS are not shown due to lack of impact at the region of interest.}
    \label{fig:supp_pdos}
\end{figure}

\begin{figure}[H]
    \centering
    \includegraphics[width=\linewidth]{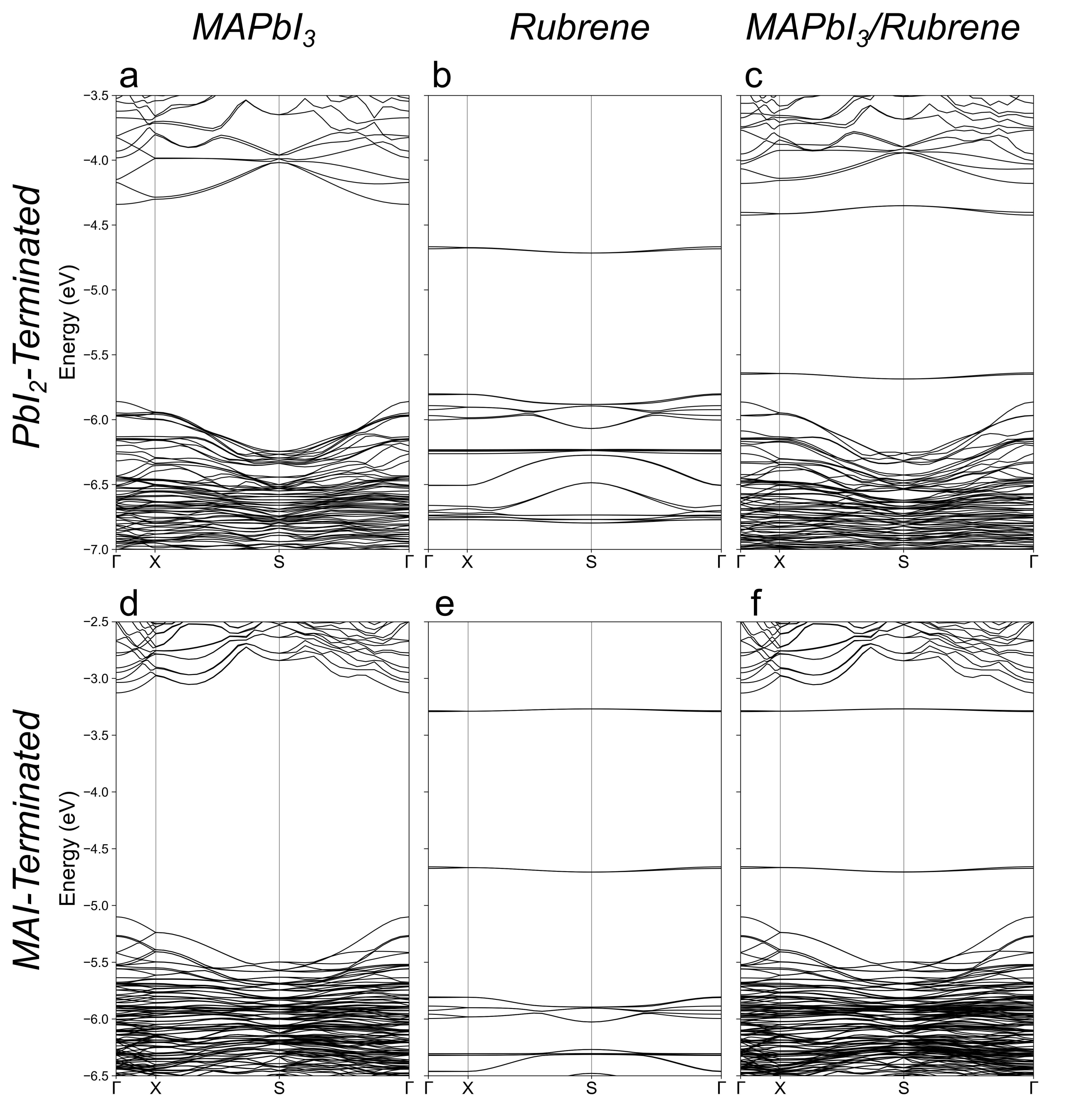}
    \caption{Band gaps of slabs and supercells for both termination types.}
    \label{fig:supp_bandgaps}
\end{figure}

\begin{figure}[H]
    \centering
    \includegraphics[width=\linewidth]{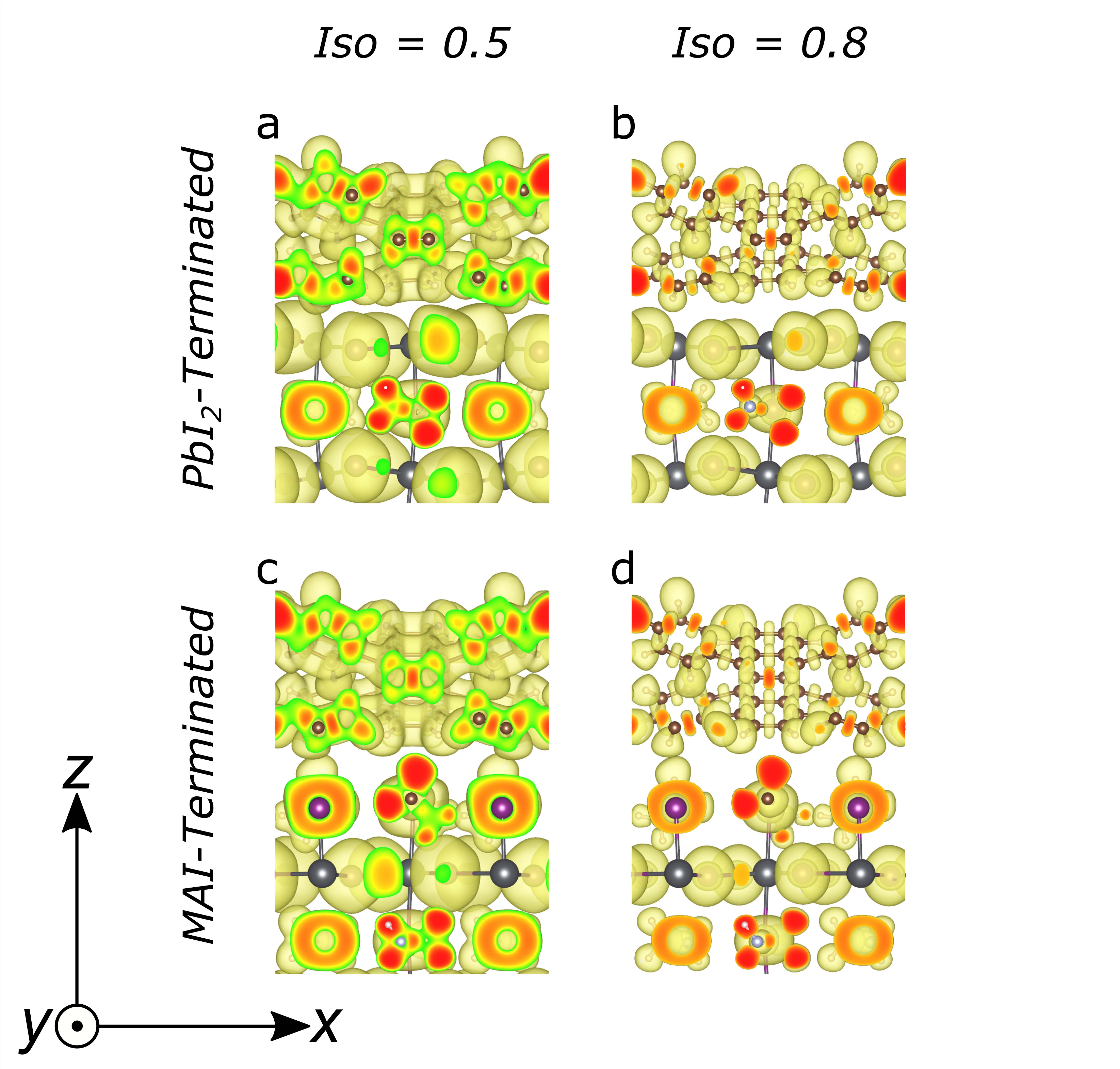}
    \caption{Electron Localisation Functions (ELF) for both (a, b) PbI$_2$- and (c,d) MAI-terminated supercells. Two isosurface values are shown in yellow with corresponding ELF isosurface values of (a,c) 0.5 and (b,d) 0.8 respectively.}
    \label{fig:supp_elf}
\end{figure}

\begin{figure}[H]
    \centering
    \includegraphics[width=\linewidth]{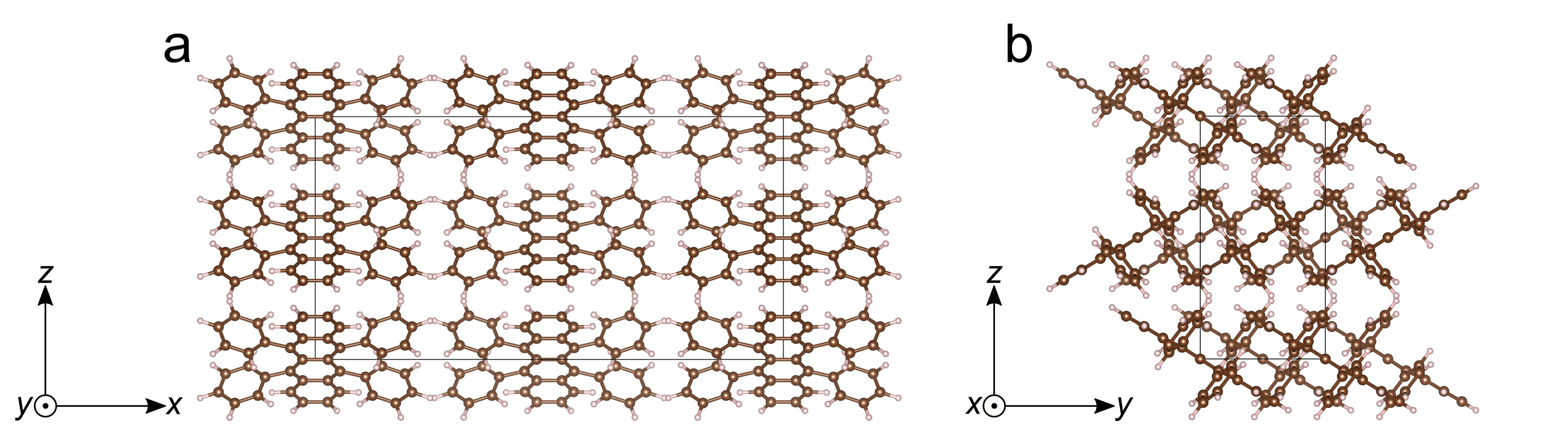}
    \caption{Relaxed geometry of the bulk rubrene supercell in the a) \textit{xz} and b) \textit{yz} planes. The box outlines the unit cell which is repeated in all directions.}
    \label{fig:supp_rubrene-geometry}
\end{figure} 

\end{suppinfo}

\end{document}